\documentclass[twocolumn]{aastex62}

\usepackage{hyperref}

\graphicspath{{./}{figures/}}

\received{...}
\revised{\today}
\accepted{...}
\submitjournal{ApJ}

\shorttitle{HD~100546}
\shortauthors{Brittain et al.}

\begin{document}

\title{High-Resolution Near Infrared Spectroscopy of HD~100546: IV. Orbiting Companion Disappears on Schedule}

\correspondingauthor{Sean Brittain}
\email{sbritt@clemson.edu}

\author[0000-0001-5638-1330]{Sean D. Brittain}
\altaffiliation{Visiting Scientist, NOAO}
\affiliation{Clemson University \\
118 Kinard Laboratory \\
Clemson, SC 29634, USA}

\author{Joan R. Najita}
\affiliation{National Optical Astronomy Observatory \\
950. N. Cherry Ave. \\
Tucson, AZ 85719, USA} 

\author{John S. Carr}
\affiliation{Department of Astronomy \\ 
University of Maryland \\
College Park, MD 20742, USA}

\begin{abstract}

HD~100546 is a Herbig Ae/Be star surrounded by a disk with a large central region that is cleared of gas and dust (i.e., an inner hole). 
High-resolution near-infrared spectroscopy reveals a rich emission spectrum of fundamental 
ro-vibrational CO emission lines whose time variable properties point to 
the presence of an orbiting companion within the hole. 
The Doppler shift and spectroastrometric signal of the CO v=1-0 P26 line,
observed from 2003 to 2013, 
are consistent with a source of excess CO emission that orbits the star  
near the inner rim of the disk. The properties of the excess emission are 
consistent with those of a circumplanetary disk. 
In this paper, we report follow up observations that confirm our earlier
prediction 
that the orbiting source of excess emission would disappear behind the 
near side of the inner rim of the outer disk in 2017.
We find that while the hotband CO lines 
remained unchanged in 2017, the v=1-0 P26 line and its spectroastrometric signal
returned to the profile observed in 
2003. 
With these new observations, we further constrain the origin of the 
emission and discuss possible ways of confirming the presence of an 
orbiting planetary companion in the inner disk.  

\end{abstract}


\section{Introduction} \label{sec:intro}
The identiﬁcation of forming planets will allow us to connect 
the initial conditions in disks to the kinds of planets that 
form, connect the initial architecture of planetary systems 
to the final distribution of planets observed around main 
sequence stars, and validate indirect signposts of gas giant 
formation. High-resolution near-infrared spectroscopy of 
the circumplanetary material surrounding forming gas giant 
planets will help to elucidate the physics of planet formation. 
For example, determining how material accretes from the 
circumstellar disk to the circumplanetary disk and finally on 
the planet will shed light on  what sets the mass of gas 
giant planets. Measuring the size and temperature of 
circumplanetary disks will provide observational constraints on 
the initial conditions for the $in~situ$ formation of satellites 
(e.g., the Gallilean moons; \citealt{Canup2002, Ward2010}).

While even the 
highest spatial resolution instruments planned for next generation facilities (e.g., the 
30-meter class telescopes) will only at best be able to marginally resolve such structures, 
high-resolution spectroscopy of warm gas using instruments on current 10-meter class 
telescopes can serve as a surrogate for high-spatial resolution. Spectroastrometric observations 
reported over the past decade, which provide 
complementary sub-milliarcsecond-scale spatial information on the dynamics of disk gas, 
demonstrate the promise of this approach 
\citep{pontoppidan2008,pontoppidan2011,brittain2009,brittain2013,brittain2014,
brittain2015}. 

Theory predicts that an orbiting giant planet will dynamically clear a gap in the 
vicinity of its orbit, or possibly deplete the entire inner disk creating a large inner hole. The large optically thin 
region that results will produce a spectral energy distribution (SED) with a large 
deficit at mid-infrared wavelengths, a transition disk SED.
Gas accreting through the disk onto the forming gas giant planet is expected to  
form a circumplanetary disk that fills roughly 1/3 of the Hill radius (e.g., \citealt{quillen1998} ). 
Recent modeling indicates that circumplanetary structures are vertically extended and 
intrinsically 3D, although they retain the rotationally supported disk found in 2D simulations 
(e.g., \citealt{Ayliffe2009, machida2010, tanigawa2012,Gressel2013, szulagyi2014-acc}).


Although observational evidence of circumplanetary structures remains 
scant, several recent studies have reported evidence for planetary 
companions in the cleared inner regions of transition disks. 
The young stars LkCa~15, PDS~70, and HD~142527 harbor companions 
(perhaps a stellar companion in the 
case of HD~142527) that appear to be accreting as evidenced by the H$\alpha$ emission 
detected from these sources \citep[][respectively]{Sallum2015,Wagner2018,Close2014}. 
Although attempts to detect a circumplanetary disk with ALMA have 
yielded little success
thus far
(e.g., \citealt{Keppler2019, Pineda2019}; but see \citealt{perez2019}, high-contrast near-infrared 
(NIR) imagery of LkCa~15 \citep{kraus2012, Sallum2015}, 
and PDS~70 \citep{keppler18, Christiaens2019} reveal compact NIR emission that could be produced by forming 
gas giant planets and/or their circumplanetary disk. 
Complementary information about any circumplanetary emission can be acquired 
with high-resolution NIR spectroscopy.


We have previously identified in the Herbig Ae/Be system HD~100546 
a possible signpost of giant planet formation: an orbiting 
source of $5\micron$ CO fundamental emission that may arise in a 
circumplanetary disk \citep{brittain2014}. Herbig Ae/Be stars are 
pre- and zero-age main sequence stars with accretion disks \citep[e.g.,][]{waters1998}.
The disk orbiting HD~100546 is devoid of dust interior to 
$R_{\rm hole}=15.7\pm0.8$\,au \citep{Jamialahmadi2018}. This region is also devoid
of molecular gas and PAHs \citep{Habart2006, brittain2009, vdp2009, carmona2011, lisk2012}

Using Phoenix on Gemini South and CRIRES on the Very Large Telescope, we inferred 
the presence of a potential companion within the inner hole, based on an excess 
component of the high-J v=1-0 ro-vibrational CO emission that varies relative 
to the stable hotband emission \citep{brittain2014}.
The variation of the Doppler shift and the spectroastrometric signal of the 
excess component 
over a 10 year period 
can
be explained by a compact source of emission in a Keplerian orbit near the
edge of the disk gap (Fig. 1). 
The discovery of a companion at the location inferred for the orbiting component has been 
reported based on high contrast near-infrared imaging \citep{currie2015}.

The CO flux 
of the excess component 
is consistent with emission from 
a circum{\it planetary} disk with a radius of $\sim$0.3~AU if we assume 
the emitting
gas is optically thick and 
at the 
same temperature as 
the circum{\it stellar} gas 
near the disk edge 
(1400~K; \citealt{brittain2013}). These values are 
consistent with the 
theoretically 
predicted thermal properties of circumplanetary disks 
of giant planets. \citet{szulagyi2014-acc} reported a disk temperature of
$\sim$2000K out to a Hill radius R$\rm_{Hill}\sim$0.8~au for the case, 
in their theoretical study, of a 
10~M$\rm_J$ planet at 5~au (see also \citealt{szulagyi2017-temp, szulagyi2017-thermo}). 
Similar temperatures have been reported for 
the inner circumplanetary disk in other three-dimensional radiation 
hydrodynamical simulations \citep{Klahr2006, Gressel2013}.


From our earlier observations, we constrained the orbital radius 
of the excess CO emission source to R=12.9$^{+1.5}_{-1.3}$~au. We were also 
able to constrain the 2003 orbital phase relative to the northwest extent of the 
semi-major axis to $\phi$=6$\degr^{+15\degr}_{-20\degr}$ \citep{brittain2014}.  
  Based on this orbit, 
we predicted that the compact emission source would 
disappear behind 
the near side of the inner disk wall sometime between February 
2017 and February 2019. Here
we report follow-up spectroscopy of HD~100546 acquired in July 2017 
and December 2017. We show that the excess emission component is no longer evident,
with the profile of the v=1-0 lines having returned to the line 
shape originally observed in 2003. 

\section{Observations} \label{sec:obs}
We acquired high-resolution (R=50,000), near-infrared spectra of HD~100546 on  
July 9, 2017 and December 3-4, 2017 using PHOENIX at the Gemini 
South telescope \citep{hinkle2003,hinkle2000,hinkle1998}. In addition
the 
hot star HIP~60718 was observed for telluric correction. 
The 4 pixel slit (0$\arcsec$.34) was used for all observations providing 1.5~km~s$^{-1}$ per pixel sampling of the spectrum. 
 The seeing in the $M$-band during these epochs ranged from 
0$\arcsec$.9--1.0$\arcsec$ 
and the airmass of the
 observations ranged from 1.5-1.6. In July, half of the spectra were 
 taken with the slit in its default position angle of 90$\degr$ east of north, and the 
 other half were taken with the slit rotated 180$\degr$. In the December, all of 
 the spectra were acquired with the slit in its default position. A 
 summary of observations is presented in Table 1.

Observations were centered near 2034~cm$^{-1}$, which covers the 
v=1-0 P26 line (hereafter we will refer to this transition as P26) and numerous hot band lines. 
These same transitions were also observed in 2003, 2006, 2010, and 2013. 
These spectra were previously presented in Brittain et al. (2014).

Observations in the $M-$band are dominated by a strong thermal background. Therefore, an ABBA nod pattern
 between two positions separated by $\sim 5\arcsec$ is used to cancel the thermal continuum to first order. The
  scans are flat fielded, cleaned of hot and dead pixels, including cosmic ray hits, and then combined in the
   sequence (A$_1$-B$_1$-B$_2$+A$_2$)/2. Because the spectra are curved along the detector, they are first
 rectified by finding the centroid of the point spread function (PSF) in each column. A third degree polynomial was 
 fit to the centroids and used to determine the shift of each column to a common row. 

In principle, the spectra can also be tilted along the slit. To check this, we compared the sky emission spectrum at 
the bottom of the slit to the spectrum at the top of the slit. The offset was less than half of a pixel along the length of 
the slit. A sky emission spectrum generated by the Spectral Synthesis Program \citep{kunde1974}, 
which  accesses the 2000HITRAN molecular database \citep{rothman2003} was used to measure the
shift of each row.  A first degree polynomial was fit to the central wavenumber of each row to measure the shift. 
Each row was then rectified in the spectral direction. 

To capture all of the CO emission, a 2$\arcsec$ window was extracted to generate the one-dimensional 
spectrum. The wavelength calibration was achieved by fitting an atmospheric transmittance model generated by 
the Spectral Synthesis Program. Each spectrum is then ratioed to a standard star observed at a similar airmass to 
remove telluric absorption lines. Areas where the transmittance is below 50\% are omitted (Figs. 2a \& 2b). 

The spectra acquired in July were observed with two PAs in order to 
make a spectroastrometric measurement of the emission lines 
\citep[see for example,][]{brittain2015, brittain18}. 
To measure the spectroastrometric signal of the emission lines, the 
centroids of the A and B rows of the rectified array are measured for 
each column and then averaged. The centroid measurements of the parallel
and anti-parallel position angles are then subtracted and divided by two
in order to remove systematic features. 
The spectroastrometric measurement is presented in figure 3. 


\section{Results}
To check for variability of the CO emission (due to artifacts and/or intrinsic variation), we averaged together six 
hotband CO lines that were not blended or significantly impacted by telluric absorption (Table 2). As in previous
analyses, the 
spectra were rescaled so that the equivalent widths of the hotband lines were 
the same from epoch to epoch 
\citep{brittain2013, brittain2014}. We then differenced the averaged profile 
from 
each epoch with 
the averaged profile from 2003 (Fig. 4). No significant variability in 
the shape of the lines is detected relative to 2003 in any of the epochs. 

A similar analysis was performed for the v=1-0 P26 CO line. The equivalent widths of the P26 line
from each epoch were scaled with the value used to scale the hotband lines from their respective epoch.
As noted in previous papers 
\citep{brittain2009,brittain2014,brittain2015}, the P26 line was substantially different both 
in flux and shape in 2006, 2010, and 2013 relative to  
2003. We have two epochs in 2017. In neither epoch is the residual of the
difference between the 2003 and 2017 epochs above the 1$\sigma$ level (Fig. 5). 
While the signal-to-noise of the spectroastrometric measurement is low due to the 
poor seeing at the time of the observation, the signal of the P26 lines is observed 
and consistent with gas orbiting in an axisymmetric disk (Fig. 3). 

Over time, the v=1-0 P26 line has varied relative to the hotband lines. Because 
these lines are observed simultaneously in the same spectrum, we are confident 
that the variability of the v=1-0 line is intrinsic to the source of emission 
rather than an artifact such as misalignment of the slit with the PSF of the 
source (e.g., \citealt{hein2014}).


It is now evident that the excess CO emission was visible for fewer than 14.5~yrs, so the 
upper limit on the period of the orbit 
is 29yrs. This allows us to refine the orbital parameters of the source of emission.
Assuming a stellar 
mass of 2.2M$_{\sun}$ \citep{Pineda2019} the object must be orbiting within 
12.3~au (Fig. 6). Adopting the 
disk 
inclination measured with mid-infrared interferometry 
(47\degr; \citealt{Jamialahmadi2018}), we also constrain the orbital phase to $\phi=-5.5\degr$ -- +8.7$\degr$ 
and the 
inner extent of the CO emission to $>$11.6~au. Thus the orbital radius of this source of 
emission is constrained to 11.6~au -- 12.3~au.  Recent ALMA measurements provide a slightly different measurement
of the 
stellar 
mass (2.05$\pm$0.01 M$_{\sun}$) and 
disk 
inclination (40.1$\degr\pm0.1\degr$; \citealt{Casassus2019}). These
values shift the center of the contours to 11.5au and -2$\degr$ so that the constraint on the orbital phase
is $\phi=-22\degr$ -- +15$\degr$ and the orbital radius is constrained to 10.5~au -- 12.3~au.
This is interior to the inner edge of the outer disk inferred from mid-infrared interferometry 
\citep[15.7 au][]{Jamialahmadi2018} and NIR high contrast imagery \citep{Follette2017}. The separation 
we infer is independent of the distance to the source as it is inferred from the velocity and 
period of the orbit, so it should not be rescaled with the updated distance to HD~100546 
available from GAIA \citep[cf.,][]{Follette2017,Jamialahmadi2018}.

\section{Discussion} \label{sec:dis}
A decade and a half of high resolution CO spectroscopy of HD~100546 provides strong 
evidence for an orbiting source of CO emission that has the temperature and emitting 
area expected for a circumplanetary disk. Some recent publications (Follette et al. 2017; Pineda et al. 2019) have
been dismissive of the CO spectroastrometric results, by making incorrect
reference to Fedele et al. (2015), whose criticism was directed at the
line profile of the ro-vibrational OH lines (see the appendix for further details). 
While it is true that an offset of the slit from the stellar position
can produce an asymmetric spectroastrometric signal and line profile 
(see \citealt{brittain2015}), this cannot explain the CO results for HD 100546.
The simple reason
is that the v=1-0 and hot-band lines were observed
simultaneously, yet they show different spectroastrometric signals and
profiles in data from 2006 to 2013. 

Even if that were not the case, slit offsets cannot produce 
the large asymmetries observed, because the slit width in our observations   
was wider than the size of the inner rim in HD~100546. Modeling of
the effects of slit misalignment on observations of axisymmetric disk
emission \citep{brittain2015} shows that, in this situation, offsets even
as large as the full slit width are unable to produce the large asymmetries
observed in the CO v=1-0 spectroastrometric signal.

Orbiting sources of emission have also been detected in the inner region 
of other transition disks, 
albeit for smaller fractions of their orbits than in the case of HD~100546. 
As one exciting example, the three emission sources in the LkCa~15 system 
show consistent orbital motion 
at $\sim 19$\,au from the star
in data taken over more than 6 years 
(Sallum et al.\ 2016, 2019 in preparation). 

Whether the orbiting source of excess CO emission in the HD~100546 system 
actually arises in a circumplanetary disk 
(or, alternatively, in an orbiting hot spot on the disk wall, or another component) 
is as yet unclear. 
Efforts to detect an associated 
source of continuum emission 
at $\sim$12~au have been inconclusive.
The separation at maximum elongation is 100~mas which is very close to the inner working angle 
of high contrast imagers on 8~m class telescopes \citep[e.g.,][]{Follette2017}. 
\citet{currie2015} reported the presence of a NIR point source interior to the inner rim 
in their image of the disk 
(HD~100546c) at a PA consistent with the inferred orbit of the CO excess emission 
source. 
Subsequent GPI imagery of the disk did not confirm this 
result \citep{Follette2017}; however, it is likely that the emission source fell
behind the coronographic mask in this epoch \citep{currie2017}. High contrast imagers 
that can more easily probe smaller orbital distances (e.g., instruments such as SCExAO on the Subaru 
Telescope and its successors) can potentially image the source of emission when 
it emerges from behind the inner rim of the disk in 2031.

If the source 
of CO excess emission 
is a circumplanetary disk, it may be detectable in the near-infrared or 
millimeter dust continuum, depending on its dust-to-gas ratio and other factors. A 
recent study of the observability of circumplanetary disks with ALMA found that a 
circumplanetary disk around a 3~M$\rm_J$ planet at 5~au would produce a flux of 
$\sim$1~mJy/beam and be detectable at 440~$\mu$m in $\sim$5~hr of integration for 
a system $\sim$100~pc away \citep{szulagyi2018}. In comparison, the ALMA upper 
limits reported by \citet{Pineda2019} at 870~$\mu$m at the position of HD~100546c  
are 5 times smaller (198 $\mu$Jy). The theoretical flux estimate assumes a dust-to-gas 
ratio of 0.01, although as the authors note, the dust-to-gas ratio could be much smaller: 
the circumplanetary disk is fed by vertical meridional flow from the outer disk, 
and only the small grains that are well coupled to the gas are entrained in the 
flow, with larger solids filtered out by the pressure bump of the outer disk. 
Perhaps the resulting low dust-to-gas ratio of the CPD reduces the submillimeter 
flux to a level consistent with the ALMA upper limit. 

There are additional lines of 
evidence suggesting that the dust-to-gas ratio 
of the accreting material in HD~100546
is substantially lower than 
0.01. 
Firstly, 
the 
stellar 
photospheric abundance of refractory material in HD~100546 is depleted relative to 
volatile elements, 
a result that led 
\citep{kama2015} 
to conclude that the dust-to-gas ratio of the accreting material is 0.001. 
Secondly, \citet{Dong2018} point out that the disk mass of HD~100546 (and other Herbig Ae/Be stars)
inferred from its submillimeter continuum emission (i.e., the dust) and the assumption that the 
dust-to-gas ratio is 0.01 implies that the accretion onto the star can only be maintained
for $\sim$10$^5$~yrs - a small fraction of the system age. Thus they conclude that the dust-to-gas 
ratio of disks around Herbig Ae/Be stars is substantially lower than 0.01 potentially making 
it difficult to detect the circumplanetary disk in the submillimeter.

Despite these considerations, \citet{perez2019} have reported the possible 
detection with ALMA of a circumplanetary disk in dust continuum emission and gas kinematics. 
The continuum emission source, located 51~mas from the star (or 5.6~au adopting a distance of 110.0$\pm$0.6~pc from \citealt{GC2}) and at a PA of 37$\degr$ in 
September 2017, is closer to the star and at a different PA than the orbiting source 
of excess CO emission studied here.

How can we confirm the presence of 
an orbiting giant planetary
companion to HD~100546 without waiting for it to emerge 
from behind the disk in 2031? 
Perhaps the most promising approach 
is through the reflex motion the companion induces in the central star. 
If a 5~M$\rm_J$ 
companion orbits 12~au from the star, 
the semi-major axis of the star's orbit is 0.24~mas.
Recent studies that have combined the Hipparcos and Gaia DR2 astrometric data 
on $\beta$~Pic have achieved a proper motion precision of 0.02--0.03~mas~yr$^{-1}$ 
\citep{Snellen2018,Dupuy2019}.

While this level of precision would be sufficient to measure the 
astrometric wobble of HD~100546, 
the $\sim 24$~year time interval between the Hipparcos and Gaia mean epochs 
is close to the inferred 27-29 year orbital period of the companion, making it difficult 
to detect a wobble with the current Gaia DR2 data. 
To detect orbital acceleration we would need to measure a difference 
in the instantaneous proper motion of the star at two epochs. 
For HD 100456, the difference between Gaia and the 24.5 year baseline is 
consistent with zero, (0.09$\pm$0.16, 0.02$\pm$0.14) mas/yr, as expected for 
the inferred orbit, although with large errors.
A better estimate will be available from the 5-year Gaia data and can be
used to test for the presence of a supra-Jovian mass companion
(T. Dupuy, private communication).

To summarize, we have presented ro-vibrational CO emission that is likely from the circumplanetary 
material orbiting a supra-Jovian mass companion in HD~100546. If confirmed, this will
be the first direct detection of emitting gas from circumplanetary material. With more
sensitive and higher resolution near and mid-infrared spectrographs on 30m class telescopes,
we will be able to study the dynamics of the circumplanetary gas in more detail and measure 
the size of disk, determine whether the gas is in a Keplerian orbit, and thus better
characterize the circumstellar disk - planet connection and the birthplace of Jovian moons.

\acknowledgments
We thank Trent Dupuy, Judit Szulagyi, Steph Sallum, and Thayne Currie for helpful discussion and advice.
This work was performed in part at the Aspen Center for Physics which is 
supported by the National Science Foundation grant PHY-1607611. Work by 
SDB was performed in part at the National Optical Astronomy Observatory. 
NOAO is operated by the Association of Universities for Research in Astronomy 
(AURA), Inc. under a cooperative agreement with the National Science Foundation. 
SDB also acknowledges support from this work by NASA Agreement No. NXX15AD94G; 
NASA Agreement No. NNX16AJ81G; and NSF-AST 1517014.

\appendix
An additional line of evidence pointing to the presence of a massive companion near the inner rim of the HD~100546 disk is the shape of the ro-vibrational OH emission line profile \citep{lisk2012, brittain2014}. The interaction between supra-Jovian mass planets and disks gives rise to a persistent eccentricity near the orbit of the planet \citep[e.g.,][]{kley06}. Emission arising from an eccentric annulus has a distinct asymmetric line profile, similar to that observed from HD100546 \citep{lisk2012}.
 
The result by \citet{lisk2012} and \citet{brittain2014} was called into question by \citet{fedele2015}. \citet{fedele2015} compared several different observations of the OH line acquired at different epochs with different slit PAs and different line profiles. Because the equivalent width of one epoch with a relatively symmetric line profile is the largest, \citet{fedele2015} conclude that this is the true shape of the OH line and that the previously reported asymmetries were spurious and caused by the occultation of the resolved disk by the relatively narrow slit.  
 
Occulting a portion of the disk emission with a narrow slit can indeed give rise to an asymmetric line profile \citep[e.g.,][]{hein2014, brittain2015, fedele2015},
although this explanation does not apply to our OH observations. Our spectra were taken in seeing-limited conditions and with a slit width large enough to  encompass the entire inner rim. The diameter of the inner hole of HD~100546 is $0\arcsec.29\pm0\arcsec.01$ \citep{Jamialahmadi2018}, whereas the slit was $\sim0\arcsec.4$,  and the seeing ranged from $0\arcsec.6-0\arcsec.8$ \citep[][]{lisk2012,brittain2014, brittain2015}. Both the seeing-limited nature of the observations and the large slit width relative to the inner rim limit the extent to which positioning offsets will affect the measured line profile.
 
In contrast, OH observations reported by \citet{fedele2015} are much more likely to suffer from a spurious line profile produced by slit occultation. Their observations were made with a narrow slit width (0$\arcsec$.2) that was very closely matched to the PSF (0$\arcsec$.17) and smaller than the diameter of the cleared region of HD~100546. Thus, it was impossible to avoid occulting some part of the inner rim at any position angle they observed. Additionally, slight offsets of the slit relative to the star could occult stellar light (or emission from the small ring of hot dust at $\sim$2~mas; \citealt{panic2014}) relative to
emission from the inner rim of the outer disk. As there is no significant NIR excess from the inner rim of the outer disk \citep{Tatulli2011, panic2014},
such a scenario would give rise to a larger line to continuum ratio for the OH emission.
 
In contrast, our observations, which reveal an asymmetric OH emission profile, were taken without adaptive optics and with a slit that encompasses the entire inner rim. Furthermore, two observations with different instruments on different telescopes (Phoenix on Gemini South and CRIRES on Gemini North) give the same line profile \citep{brittain2014}. Observations of other transition disks with emission arising from the outer disk using a similar instrumental set up also do not reveal such line asymmetries \citep[e.g.,][]{brittain2007,brittain2009,lewis2010,brittain18}. Thus we conclude that the OH line profiles of HD~100546 are indeed intrinsically asymmetric and consistent with emission arising from an eccentric annulus.

\bibliographystyle{apj}
\bibliography{cite.bib}	

\begin{deluxetable*}{ccccc}
\tablecaption{Journal of Observations}
\tablewidth{0pt}
\tablehead{ \colhead{Date}	&	\colhead{Star}	&	\colhead{Int Time}	&	\colhead{Seeing}	&	\colhead{SNR}	\\
	      {}                &		    {}      &		     {}         &	   {}                 &	{}	\\ }
\startdata
July 09, 2017	    &	$\alpha$ Cru	&	8m	&	\nodata	        &	\nodata	\\
	                &	HD~100546	    &	88m	&	0$\arcsec$.9	&	120	\\
December 03, 2017	& 	HIP~60718	    &	12m	& 	\nodata	        &	\nodata	\\
	                &	HD~100546	    &  	40m	&	1$\arcsec$.0	&	\nodata	\\
December 04, 2017	& 	HIP~60718	    &	12m	&	\nodata	        &	\nodata	\\
	                & 	HD~100546	    &	64m	&	0$\arcsec$.8	&	130\tablenotemark{a}	\\
\enddata
\tablenotetext{a}{The signal-to-noise ratio of the combined data from December 3 and December 4.}
\label{tab:t1}
\end{deluxetable*}

\begin{deluxetable*}{cccccccc}
\tablecaption{Hotband and $^{13}$CO lines}
\tablewidth{0pt}
\tablehead{ \colhead{Isotopologue} & \colhead{v$^{\prime\prime}$}	&	\colhead{J$^{\prime\prime}$}	&	\colhead{v$^{\prime}$}	&	\colhead{J$^{\prime}$}	&	\colhead{E$^{\prime\prime}$}	&	\colhead{$\tilde\nu$}	&	\colhead{A}	\\
 {} & {} & {} & {} & \colhead{cm$^{-1}$} & \colhead{cm$^{-1}$} & \colhead{s$^{-1}$} \\ }
\startdata
$^{12}$CO & 1	&	21	&	2	&	20	&	4103.67	&	2029.66	&	30.28	\\
$^{12}$CO & 2	&	15	&	3	&	14	&	5794.30	&	2030.16	&	45.75	\\
$^{12}$CO & 1	&	10	&	2	&	9	&	3353.92	&	2032.87	&	30.55	\\
$^{12}$CO & 3	&	8	&	4	&	7	&	7566.63	&	2033.14	&	62.96	\\
$^{13}$CO & 0	&	16	&	1	&	15	&	1557.06	&	2033.42	&	15.03	\\
$^{12}$CO & 2	&	14	&	3	&	13	&	5737.76	&	2034.41	&	46.15	\\
\enddata
\label{tab:t2}
\tablecomments{The hotband lines and the $^{13}$CO lines are excited by UV fluorescence \citep{brittain2009}.}
\end{deluxetable*}

\begin{figure*}
    \centering
    \includegraphics[width=1\textwidth, trim={0.5in 4in 0.5in 0.5in},clip]{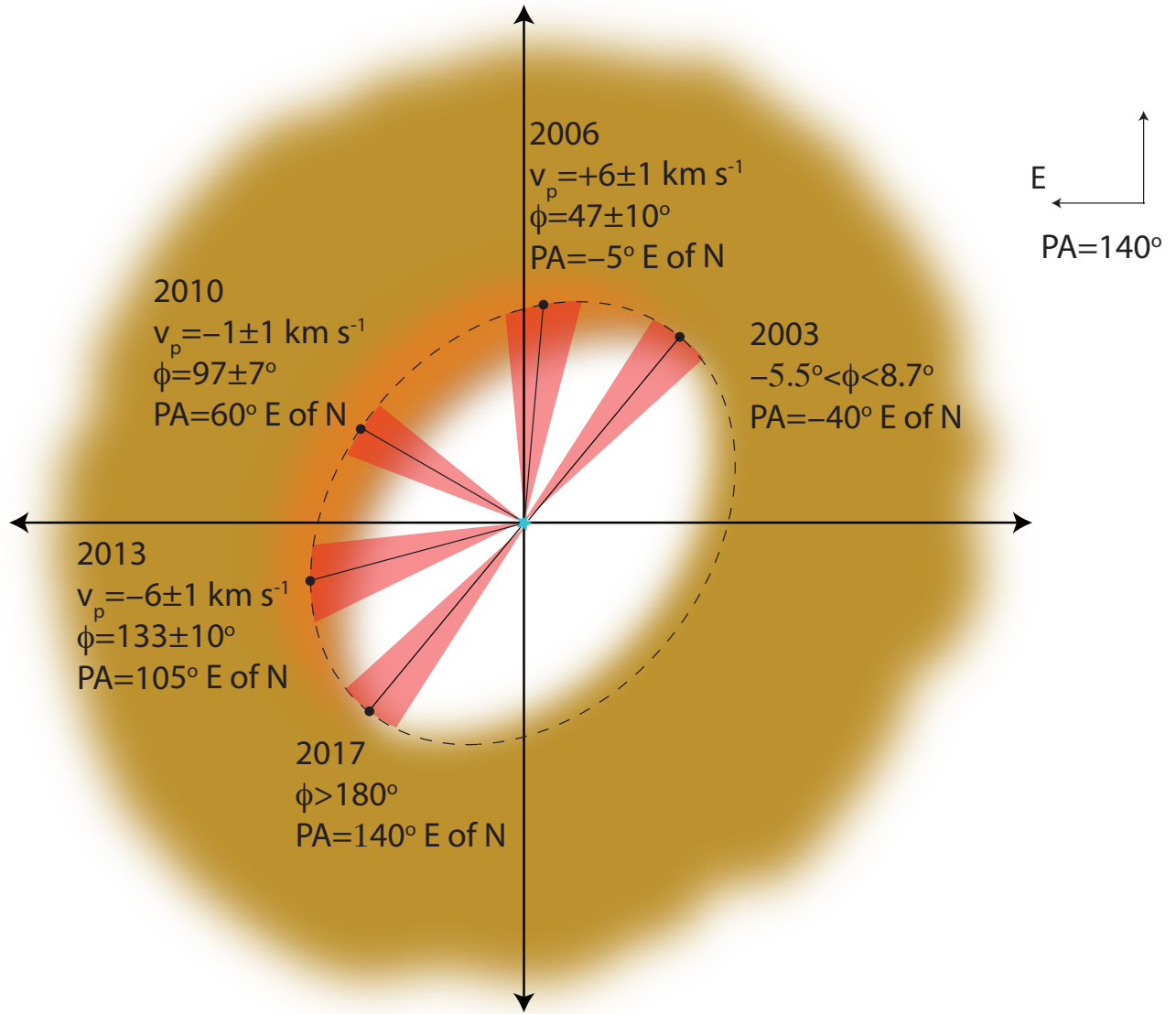}
    \caption{Schematic of the disk and orbiting source of excess CO emission identified in data taken from 2003 to 2017. The orbit is represented by the black
dashed line. The disk wall of the disk is shaded 
orange. The location of the source of the emission excess is labeled with a black
dot, and the uncertainty in the phase of the orbit is
represented by the red sectors. We assume the 
excess CO emission is hidden by the near side of the circumstellar
disk in 2003 and 2017. }
    \label{fig:1}
\end{figure*}

\begin{figure*}
        \includegraphics[width=1\textwidth,trim={0.0in 1.2in 0.0in 1.8in},clip]{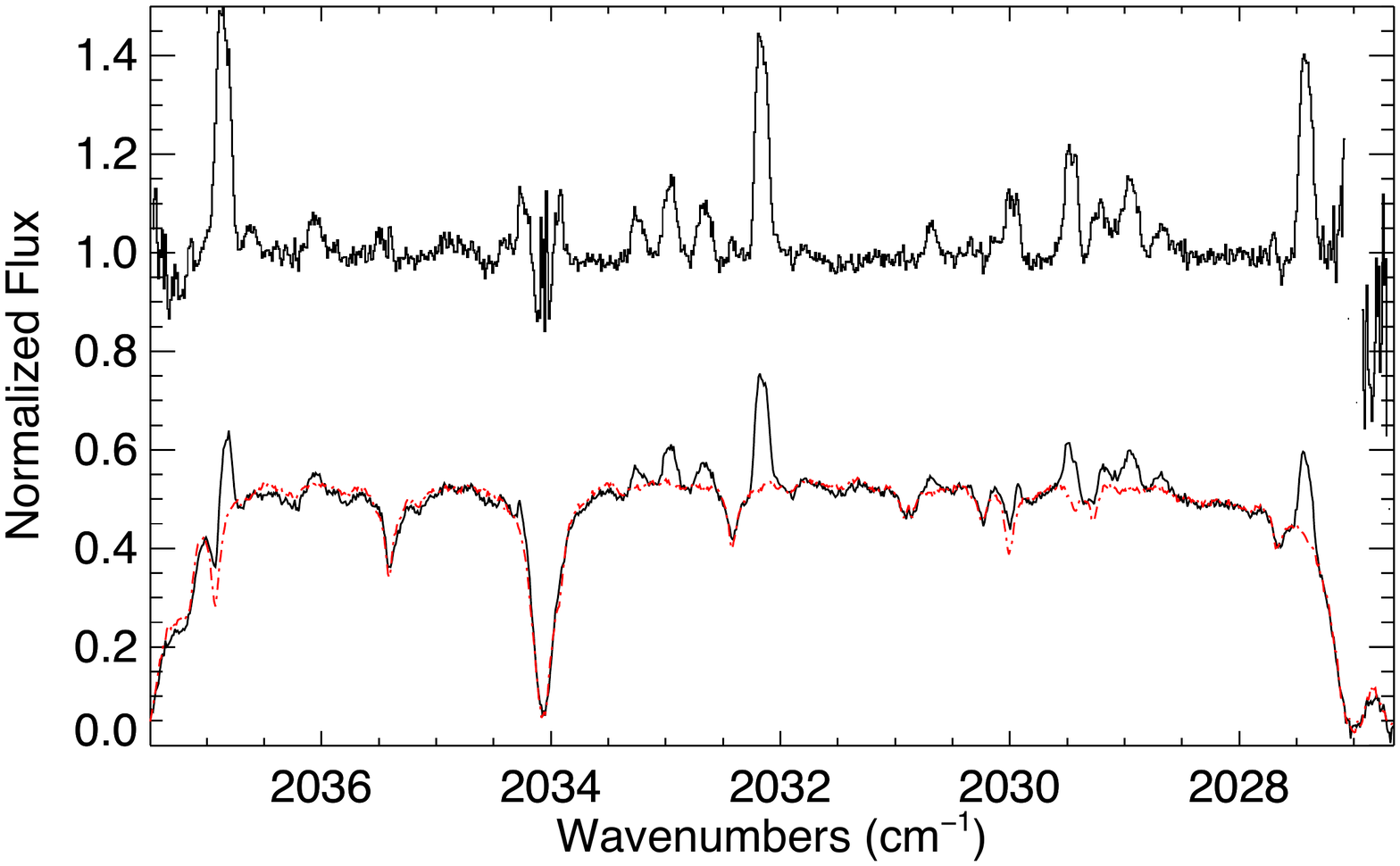}
        \label{fig:2a}

        \includegraphics[width=1\textwidth,trim={0.0in 1.2in 0.0in 1.8in},clip]{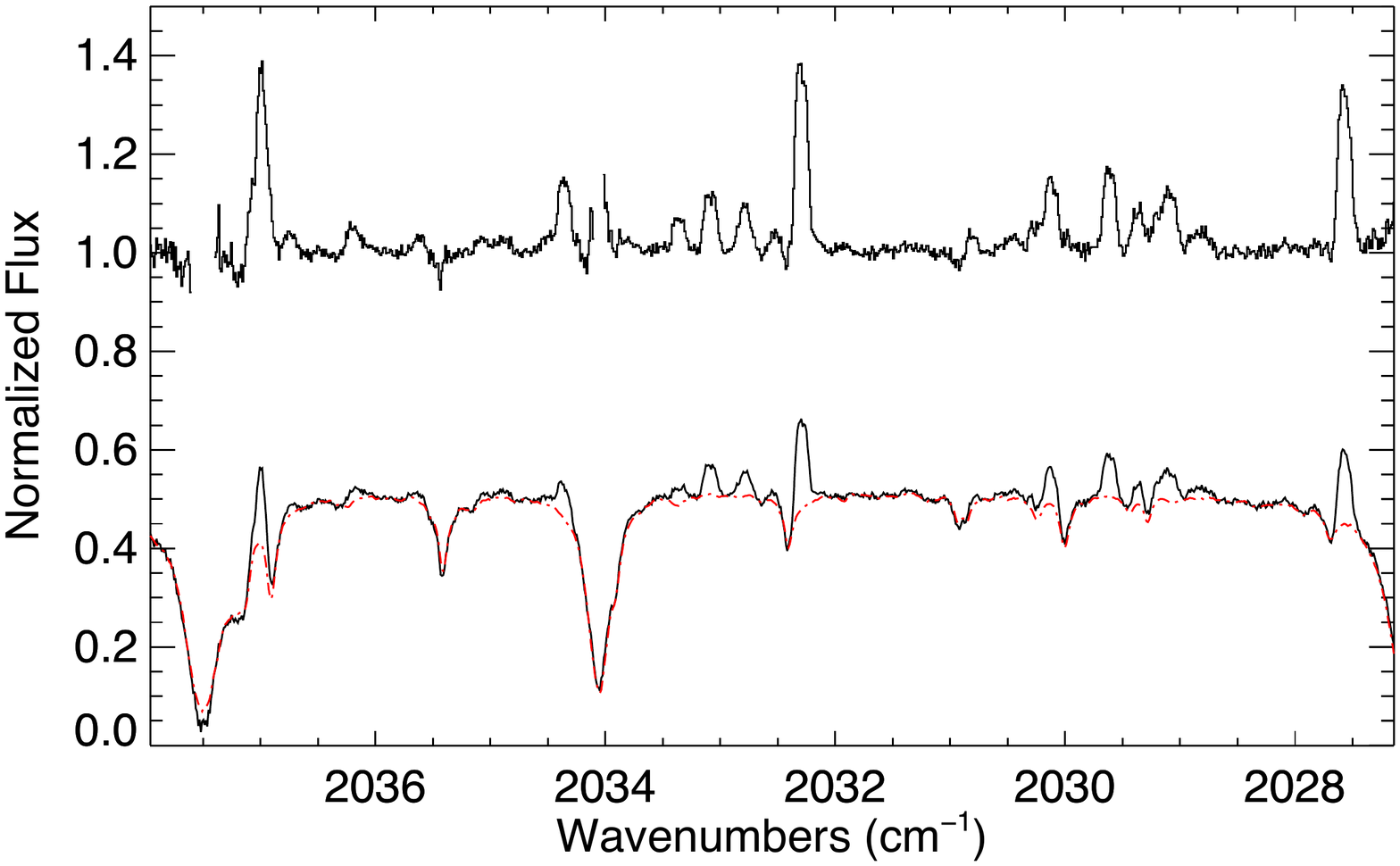}
        \label{fig:2b}
 \caption{$M-$band spectrum of HD~100546 in July 2017 (upper panel) and December 2017 
 (lower panel). The spectrum of HD~100546 is plotted black with the telluric standard 
 plotted in red. The ratioed spectrum is plotted above. Gaps in the spectra are regions 
 where the atmospheric transmittance is less than 20\%.   }
 \label{fig:fig2}

\end{figure*}

\begin{figure*}
\includegraphics[width=0.8\textwidth,trim={0.0in 0in 0.0in 0in},clip]{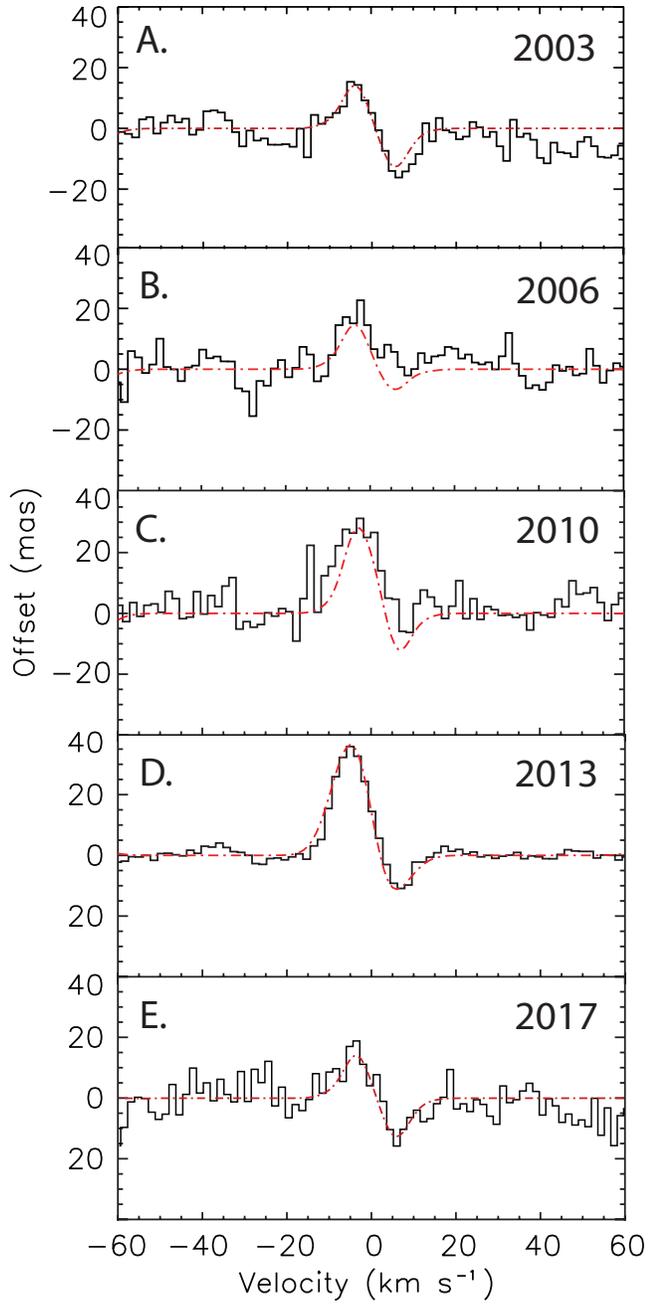}
        \caption{Spectroastrometric signal of the P26 line observed in each epoch (2003-2017). The spectroastrometric 
        signal of the P26 line is detected in 2017 and consistent with gas in an axisymmetric, differentially 
        rotating disk. It is also consistent with the signal observed in 2003. The asymmetry observed in previous epochs
        (2006 -- 2013), due to the excess emission, is no longer observed.}
        \label{fig:3}
\end{figure*}

\begin{figure*}
\includegraphics[width=1\textwidth,trim={0.5in 2.2in 0.5in 2in},clip]{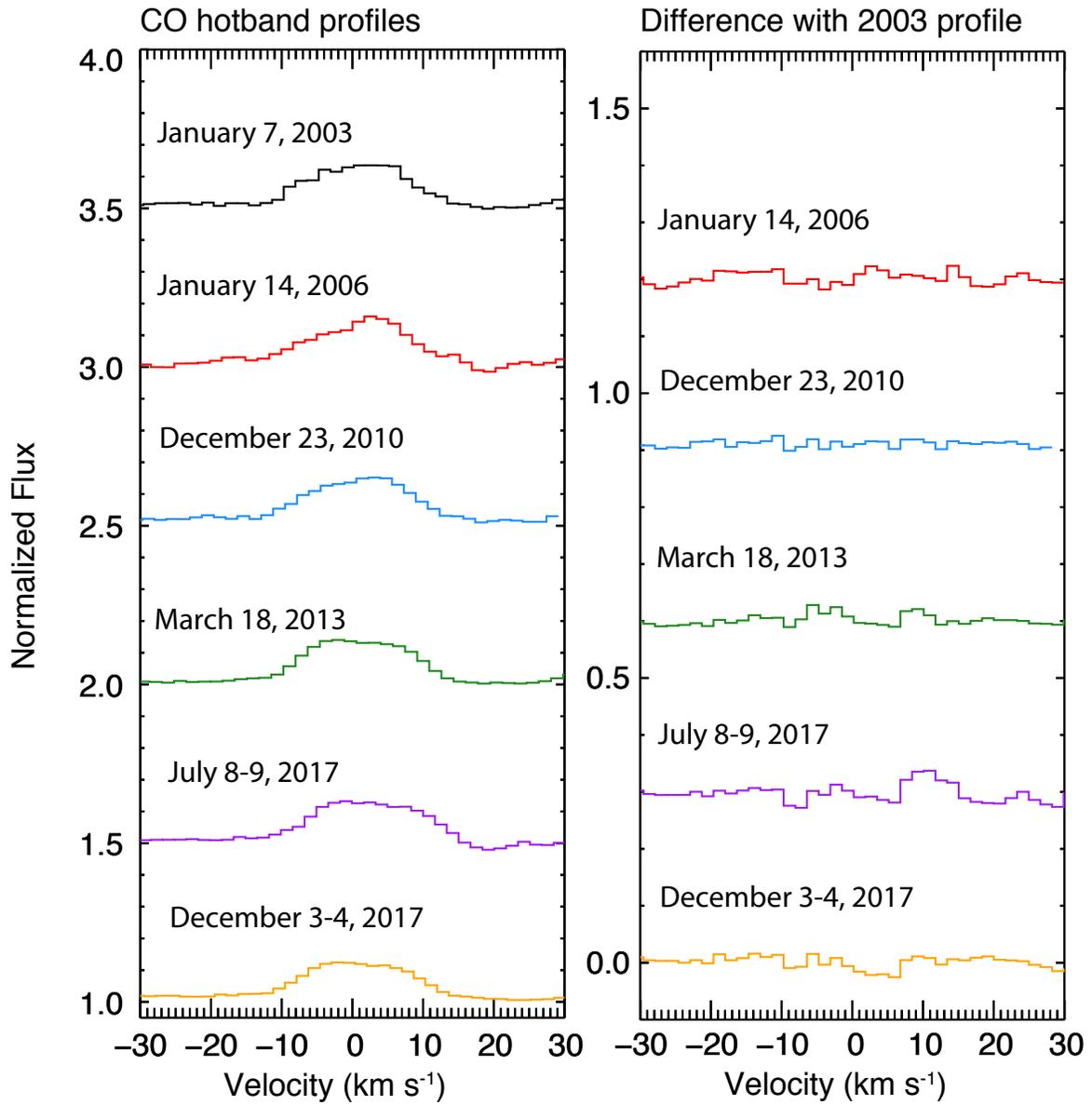}
        \caption{Multi-epoch observations of the average hotband line profile. 
        Six hot band CO lines are averaged together and plotted for each date 
        of observation (left panel). Each profile is differenced with the profile
        observed on January 7, 2003 (right panel). At the level of the S/N of the observations, 
        there is no evidence of variability of the hotband CO lines.}
        \label{fig:4}
\end{figure*}

\begin{figure*}
\includegraphics[width=1\textwidth,trim={0.5in 2.2in 0.5in 2in},clip]{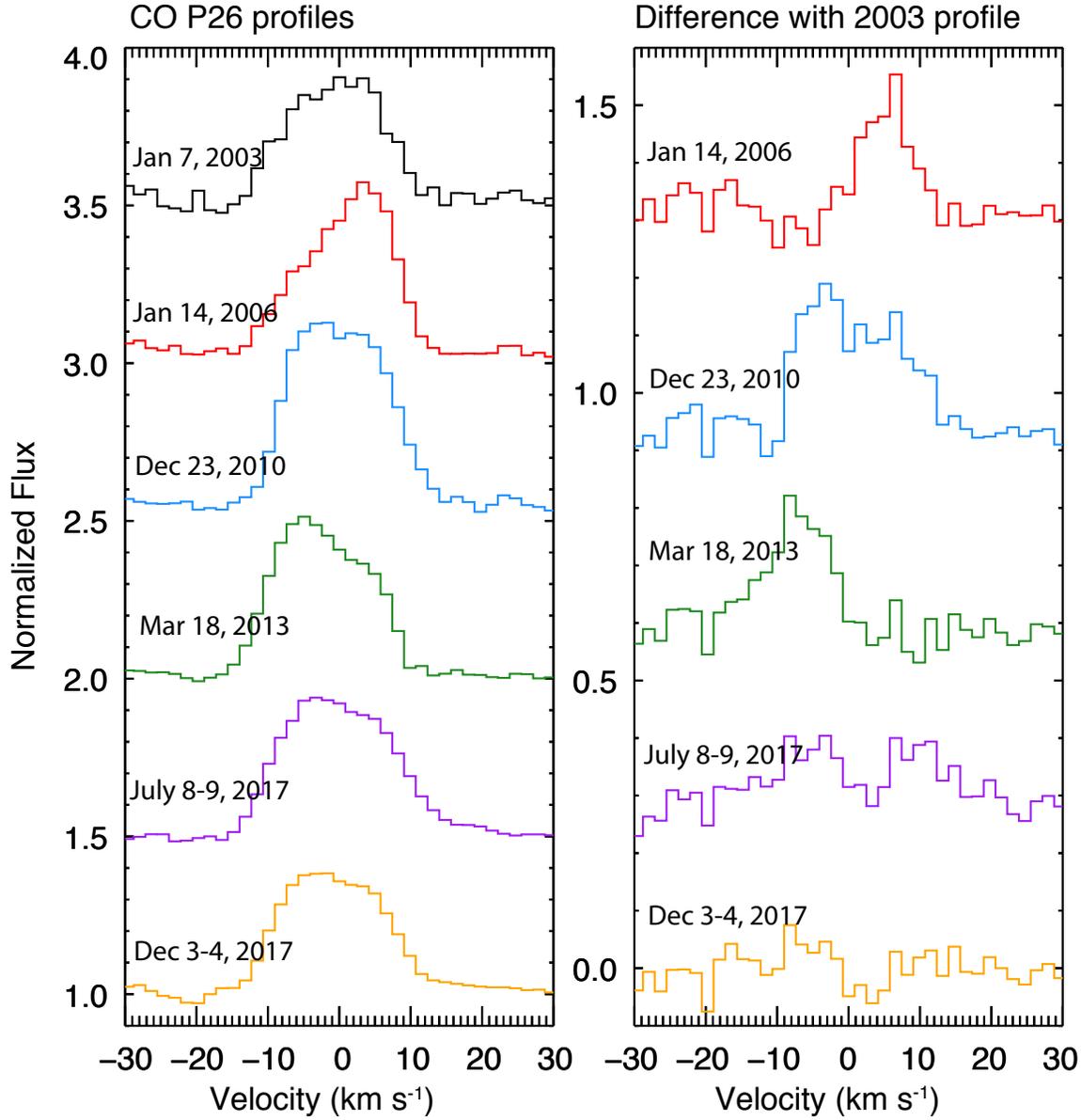}
        \caption{Multi-epoch observations of the v=1-0 P26 line profile. 
        The v=1-0 P26 line is plotted over six epochs (left panel). 
        Each profile is differenced with the profile
        observed on January 7, 2003 (right panel). Unlike the hot band lines, there is clear evidence
        of variability. The excess emission shifts from +6 km s$^{-1}$ on January 14,
        2006 to --6 km s$^{-1}$ on March 18, 2013. By July 2017, the excess emission
        is gone.}
        \label{fig:5}
\end{figure*}

\begin{figure*}
    \centering
    \includegraphics[width=1\textwidth]{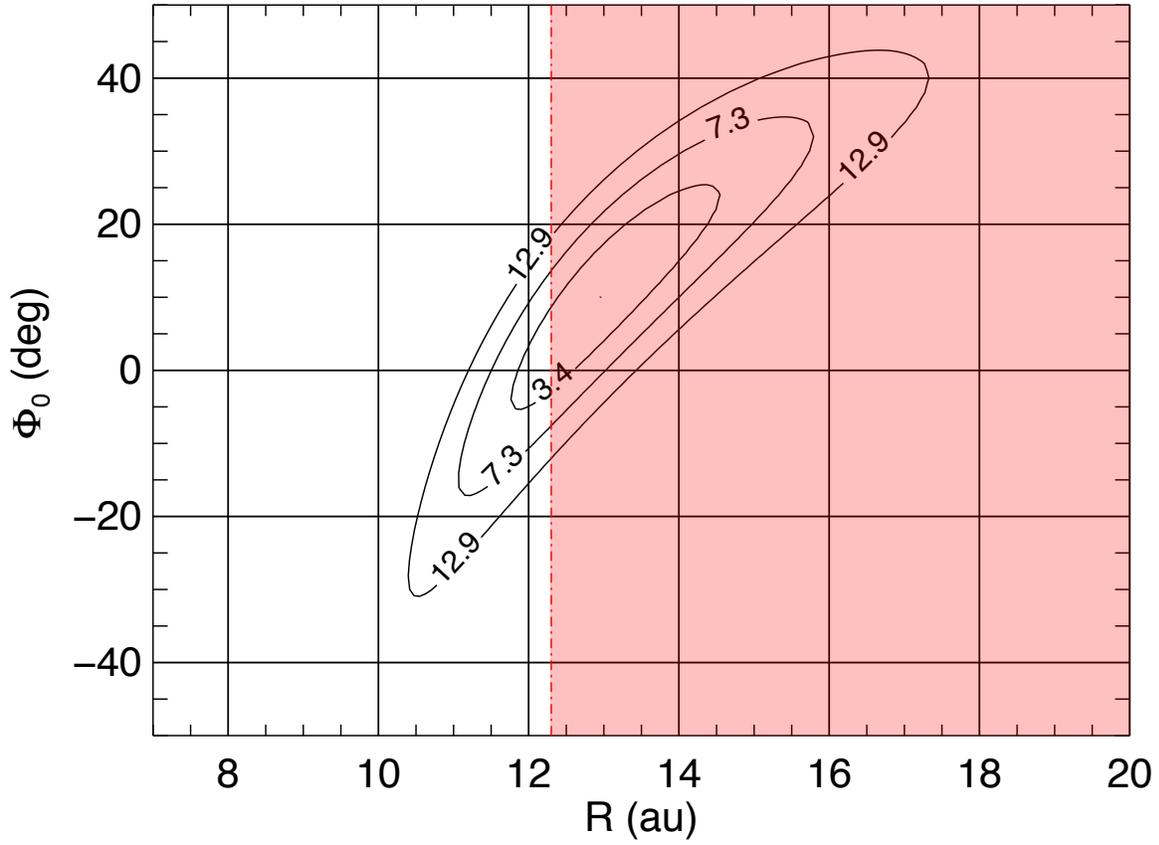}
    \caption{Constraint on the orbital radius R and orbital phase in 2003 ($\phi_0$). The constraint is placed by 
    fitting the projected velocities and relative phases in 2006, 2010, and 2013, assuming 
    the excess CO emission is in a circular Keplerian orbit, the stellar mass is 2.2M$_{\sun}$, 
    and the inclination is 47\degr. The 1$\sigma$ , 2$\sigma$ , and 3$\sigma$ confidence 
    intervals are plotted. The disappearance of the emission by July 2017 indicates that 
    the orbital period must be less than 29 years, thus we can rule out solutions with 
    R$>$12.3~au (shaded region). Thus we find find that 11.6$<$R$<$12.3~au and 
    $-5.5\degr<\phi_0<8.7\degr $. }
    \label{fig:6}
\end{figure*}

\end{document}